# Adaptive Depth Imaging with Single-Photon Detectors

Weiji He,[1,*] Zhenchao Feng,[1] Jie Lin,[1] Shanshan Shen,[1] Qian Chen,[1] Guohua Gu[1]

Beibei Zhou[2] Ping Zhang[3]

[1]Jiangsu Key Laboratory of Spectral Imaging & Intelligence Sense (SIIS), Nanjing University of Science and Technology, Nanjing, China, 210094

[2]Nanjing University of Science and Technology, Nanjing, China, 210094

[3]Jiangsu North Huguang Optics & Electronics Co.,Ltd, Wuxi, China, 214000

[*]Corresponding author: hewj@mail.njust.edu.cn

*Abstract*— For active optical imaging, the use of single-photon detectors can greatly improve the detection sensitivity of the system. However, the traditional maximum-likelihood based imaging method needs a long acquisition time to capture clear three-dimensional (3D) image in low light-level. To tackle this problem, we present a novel imaging method for depth estimate, which can obtain the accurate 3D image in a short acquisition time. Our method combines the photon-count statistics with the temporal correlations of the reflected signal. According to the characteristics of the target surface, including the surface reflectivity, our method is capable of adaptively changing the dwell time in each pixel. The experimental results demonstrate that the proposed method can fast obtain the accurate depth image despite the existence of strong background noise.

*Index Terms*— adaptive depth imaging, temporal correlations, single-photon detectors, photon-counting LIDAR

## I. INTRODUCTION

Improvements in three-dimensional imaging and single-photon detection technology, which employ depth and reflectivity imaging approach, have facilitated many emerging applications, such as, machine vision, industrial model design, and atmospheric remote sensing [1-3]. Using the 3D imaging light detection and ranging system (LIDAR) [4-7], with a Geiger-mode APD (Avalanche Photo Diode) as the single-photon detector, we can accurately acquire the depth and reflectivity images of target. Traditionally, the depth estimate based on maximum likelihood generates the photon counting histogram by a long dwell time to acquire the depth and reflectivity estimation of target, which respectively corresponds to the peak location and the amplitude of the photon-count histogram. It requires $10^3 \sim 10^6$ photons to accurately estimate the depth and reflectivity information of target. In addition, the target with low reflectivity needs much longer dwell time to mitigate the effect of Poisson noise. Therefore, imaging methods with fixed dwell time [8, 9] lead to either under-sampling or long acquisition time when dealing with the target without any prior knowledge in low light-level.

Hyunjung Shim and SeungKyu Lee [10] used a hybrid exposure technique to improve the severely noisy depth estimation caused by the target reflectivity and the limitation of the detector dynamic range. For this imaging method, it needed to take twice detections of long and short exposure time to determine the best depth estimation of each pixel. Ahmed Kirmani and Dheera Venkatram et al. [11-13] brought out a mathematical probability model of photon counting process equipped with the single-photon detector. They used the first detected photon to estimate the depth and reflectivity images of target combining the spatial correlation among adjacent pixels. However, the first detected photon may be a noise photon especially near the edges of target, which will seriously affect the estimation accuracy.

In this paper, an adaptive depth imaging strategy (ADI) is derived to solve the problem of under-sampling or long acquisition time. Photons generated by the ambient light and dark count are defined as noise photons; photons generated by the reflected laser light are defined as signal photons. The TOFs (the plural of "time-of-flight") of noise photons are independently and uniformly distributed in the whole detection period; however, because the TOFs of signal photons are characterized by the time-shifted laser pulse shape [20], they are mainly and correlatively distributed in the range of the full width half maximum (FWHM) of the emitted laser pulse. Based on this observation, ADI distinguishes signal photons from noise photons and then iteratively updates the depth estimation of target. Since only signal photons contribute to

ADI, the accuracy of depth estimation is improved. Comparing to the conventional depth estimate methods, ADI need not to generate the photon-count histogram. Furthermore, according to the characteristics of the target surface, including the surface reflectivity, ADI is capable of adaptively changing the dwell time in each pixel. We experimentally demonstrated that this method can quickly obtain a clear depth image of target in the presence of high background noise. The rest of paper is organized as follows: the image theory based on probability statistics is described in detail in Section 2. In Section 3, we provide a series of experiments to analyze the imaging performance and the effect of parameters. In Section 4, we briefly conclude our work.

## II. Theory

A schematic diagram of the LIDAR system is shown in Figure 1 [14]. A periodically pulsed laser source is used to trigger a PicoHarp 300 TCSPC (time corrected single photon counting) module (4 ps minimum time-bin width). The laser pulse is a Gaussian waveform $s(t)$, which has a repetition period of approximately $T_r = 400\ ns$ and a full width half maximum (FWHM) of $T_p = 250\ ps$ at a wavelength of 830 nm. The laser is deflected in two spatial dimensions of X/Y using a pair of computer-controlled galvanometer mirrors. The Geiger-mode APD has a dead time of 50ns and dark count of less than 100 counts per second.

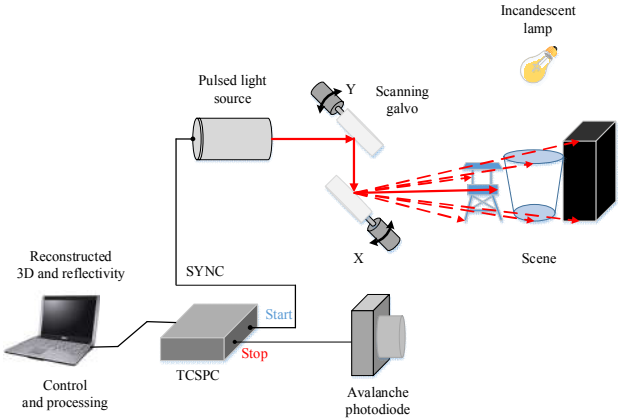

Fig. 1. The experimental system.

The spatial location of each pixel in target is marked as $(i, j, z_{i,j})$ $i, j = 1, 2, ..., N$. The distance between the target and the detector is denoted as $z_{i,j}$ (herein the plane of light source is used as the reference plane). Considering the effect of distance aliasing [15, 16], we assume $T_r > 2z_{max}/c$, where $T_r$ is the pulse-repetition period of laser, $z_{max}$ is the maximal distance, and $c$ is the speed of light. Generally, because the FWHM of laser pulse $T_p$ decides the achievable depth resolution without the ambient light [23], we assume $T_p \ll 2z_{max}/c$. The measurement time (the gate) is defined as the particular time duration that the APD is biased into Geiger-mode [18], which is set to 200 ns in our experiments.

Then the measurement time $t_M$ is divided into $s = t_M/\Delta = 200/T_p = 800$ time-bins [7], where the width (time duration) of each time-bin $\Delta$ is set to $T_p$ (the FWHM of laser pulse). The TOF (time-of-flight) $t_{i,j}$ and the time-bin number (sequence number) of the detected photons are provided by the system electronics.

In our experiments, we use root mean-square error (RMSE) to quantify the recovery accuracy of depth estimation, which is defined as

$$RMSE = \sqrt{\frac{1}{m^2}\sum_{i=1}^{m}\sum_{j=1}^{m}(z_{i,j} - z'_{i,j})^2} \quad (1)$$

where $z_{i,j}$ is the measured target depth, $z'_{i,j}$ is the real target depth at pixel $(i, j)$, and $m$ is the size of the depth image. Meanwhile, we use the signal-to-background ratio (SBR) to quantify the effect of background noise, which is defined as

$$SBR = \frac{N_{signal}}{N_{noise}} = \frac{N_{total} - N_{noise}}{N_{noise}} \quad (2)$$

where $N_{total}$ is the number of the total photon counts in the measurement time, $N_{noise}$ is the number of the photon counts generated by background noise and dark count, and $N_{signal}$ is the number of the photon counts generated by the reflected laser light.

### A. Strategy

Under the existence of background noise, the RMSE in first-photon estimate is $(c/2)\sqrt{\left(T_p^2 + T_r^2/12\right)/2}$ [11], which is not depend on whether this photon is due to noise or the reflected signal. Thus in first-photon imaging [11], only using the TOF of a single photon easily appears an estimation error especially near the edges of target. In our method, we use the TOFs of several photons combining the temporal correlations of signal photons to increase the reliability of depth estimate and distinguish signal photons from noise photons.

The adjacent $n$ photons in the time-line are taken as an elementary unit. We take an elementary unit of $n = 3$ in our experiments. $\{t_{i,j}^{(l)}\}_{l=1}^{n}$ is the TOF dataset of a photon unit. The absolute TOF differences (sorted in ascending order) are:

$$\left|t_{i,j}^{(1)} - t_{i,j}^{(2)}\right|, \left|t_{i,j}^{(2)} - t_{i,j}^{(3)}\right|, ... \left|t_{i,j}^{(n)} - t_{i,j}^{(n-1)}\right|. \quad (3)$$

Generally, for a Geiger-mode APD the probability of detection follows Poisson statistics [17, 18]. $S_r(t)$ is defined as the rate function for the number of the photon counts generated by the reflected laser light. $B = \eta_d b + d$ is defined as the rate function for the number of the photon counts generated by background light ($\eta_d b$, where $\eta_d$ is the quantum efficiency of the detector) and dark count ($d$), which is assumed to be constant and straightforward to be measured. Taking no account of the multiple reflected laser pulses, the rate function for the total photon counts is $\lambda(t) = S_r(t) + B$. On the basis of the Poisson statistics, the detection probability of photons in the $r-th$ time-bin is [17, 18]

$$P_d(r) = \exp[-\int_0^{(r-1)\Delta} \lambda(t)dt] \times \left\{1 - \exp[-\int_{(r-1)\Delta}^{r\Delta} \lambda(t)dt]\right\} \quad (4)$$

where $\Delta$ is the width of a time-bin.

Since Equation (4) is hard to be computed in practice, we assume $S_r(t)$ is constant in a small time duration of $[t_1, t_2]$ for simplification, where $t_1 = \min\{t_{i,j}^{(l)}\}_{l=1}^n$ and $t_2 = \max\{t_{i,j}^{(l)}\}_{l=1}^n$ in the TOF dataset $\{t_{i,j}^{(l)}\}_{l=1}^n$ of an elementary unit. The mean number of the photon counts generated by the reflected laser light in $[t_1, t_2]$ is approximately calculated from the laser radar equation [21]

$$S_r = S_r(t) \times (t_2 - t_1) = \frac{E_T \lambda}{hc} \eta_d \left(\frac{FOV}{\theta_T}\right)^2 \frac{\rho}{\pi} COS\theta_{t\arg et} \frac{A_R}{r^2} \eta_T \eta_R \eta_A^2. \quad (5)$$

In our experiments, the variables in Equation (5) are measured as follows: the wavelength of laser pulse is $\lambda$=830 nm; $E_T$ is the transmitted energy of the laser pulse; the Plank constant is $h = 6.63 \times 10^{-34} \; m^2kg/s$; the speed of light is $c = 3 \times 10^8 \; m/s^2$; the quantum efficiency of the detector is $\eta_d = 0.35$; the Field of View (FOV) of the receiving optics is 0.4 mrad; the divergence angle of the laser beam $\theta_T$ is 1.33 mrad; the incidence angle between the laser beam and the surface normal is $0° \leq \theta_{t\arg et} \leq 10°$, and $COS\theta_{t\arg et}$ is approximately equal to 1; the aperture of the detector is $A_R = \pi(85)^2 \; um^2$; The distance between the target and the detector is $r$=20 m; the efficiency of the transmitting optics is $\eta_T$=80%; the efficiency of the receiving optics is $\eta_R$=80%; the reflectance of target $\rho$ is approximately estimated by $\rho = N_{i,j}/x_{i,j}$ [22], where $x_{i,j}$ is the number of the transmitted laser pulses and $N_{i,j}$ is the number of the photon counts at pixel $(i, j)$; the atmospheric transmission factor $\eta_A$ is 1. Therefore, the detection probability of signal photons in an elementary unit is [17, 18]

$$P_{d_{i,j}} = \exp[-Bt_1] \times [1 - \exp(-S_r - B(t_2 - t_1))]. \quad (6)$$

At pixel $(i, j)$, the Rank-Ordered Absolute Difference ($ROAD_{i,j}$) is defined as the sum of the first $n/2$ absolute differences from the collection in Equation (3) [19]. Because the TOFs of signal photons are characterized by the time-shifted laser pulse shape [20], signal photons are mainly and correlatively distributed in the range of $T_p$. Meanwhile, the TOFs of noise photons are uniformly and independently distributed in the pulse repetition period $[0, T_r)$. Thus, a binary hypothesis test is applied to distinguish signal unit from noise unit:

$$\begin{aligned} &if \; ROAD_{i,j} > P_{d_{i,j}} \; n/2 \, T_p, \; then \; the \, detected \, photon \; \text{unit} \; is \, noise \, unit \\ &if \; ROAD_{i,j} \leq P_{d_{i,j}} \; n/2 \, T_p, \; then \; the \, detected \, photon \; \text{unit} \; is \, signal \, unit. \end{aligned} \quad (7)$$

If the photon unit is judged as a noise unit, ADI continues to compute the $ROAD_{i,j}$ of next photon unit until the first signal unit is found. We use the binary hypothesis test to reject noise photons and only keep signal photons for further processing. ADI uses the mean value of the TOF dataset of the first detected signal unit as the initial depth estimation:

$$T_{i,j}^{(0)} = \frac{1}{n} \sum_{l=1}^n t_{i,j}^{(l)}. \quad (8)$$

Considering the effect of the reflected laser-pulse broadening, the initial depth estimation of Equation (8) is likely to occur in the following three intervals: 1) If it occurs in the rising edge of the reflected laser pulse, the final depth estimation will be small; 2) If it occurs in the falling edge of the reflected laser pulse, the final depth estimation will be large; 3) If it occurs in the middle of the reflected laser pulse, the final depth estimation will be ideal. Therefore, the initial depth estimation is used to seriatim test the subsequent detected photon. If the TOF of the subsequent detected photon $t_{i,j}^{(\zeta)}$ meets the requirement of Equation (9), this photon is regard as a signal photon.

$$\left|t_{i,j}^{(\zeta)} - T_{i,j}^{(\zeta-1)}\right| \leq T_p, \; \zeta = 1, 2, 3, \ldots, K \quad (9)$$

where $K$ is the threshold. Then the TOF of this photon is used to update the depth estimation of target in real time:

$$T_{i,j}^{(\zeta)} = \left(T_{i,j}^{(\zeta-1)} + t_{i,j}^{(\zeta)}\right)/2, \; \zeta = 1, 2, 3, \ldots, K \quad (10)$$

With the value of $T_{i,j}^{(\zeta)}$ updating, the photons that meet the requirement of Equation (9) are converged in the range of $T_p$. According to the TOF distribution feature of signal photons, those photons are more likely to be signal photons. Because more and more identified signal photons are used to update the value of $T_{i,j}^{(\zeta)}$, this possibility is increased with the depth estimation updating. After the depth estimation iteratively updating $K$ times ($K$ is the threshold), we get the TOF dataset of signal photons $\{t_{i,j}^{(\zeta)}\}_{\zeta=1}^K$. Therefore, the depth estimation of target is

$$z_{i,j}^{Ada} = (c/2)T_{i,j}^{(K)} = (c/2)\sum_{\zeta=1}^K p^{(\zeta)}t_{i,j}^{(\zeta)} \quad (11)$$

where $p^{(\zeta)}$ is the weight of $t_{i,j}^{(\zeta)}$ and satisfies $p^{(1)} < p^{(2)} < \cdots < p^{(K)}, \sum_{\zeta=1}^K p^\zeta = 1$.

Generally, the number of the transmitted laser pulses decides the dwell time in this pixel. ADI calculates the minimum dwell time in each pixel by counting the minimum number of the transmitted laser pulse until achieving the threshold $K$. Then in ADI the minimum number of the transmitted laser pulse is different since the photon counts of each pixel are different. Therefore, the minimum acquisition time in each pixel is obtained, and the useless acquisition time is removed.

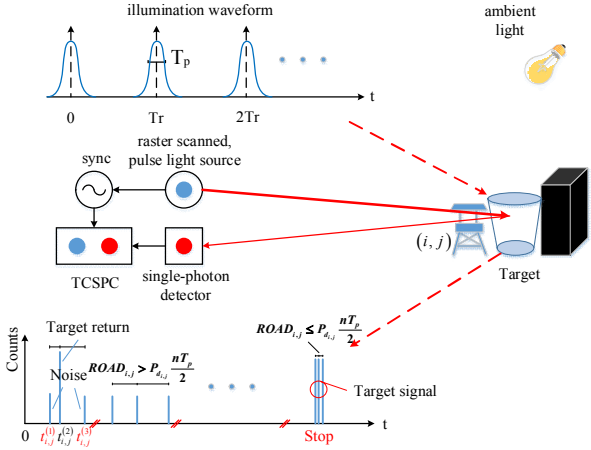

Fig. 2. (a) is the schematic diagram of the adaptive depth imaging (ADI) strategy. The adjacent three photons in the time-line are taken as an elementary unit. $t_{i,j}^{(1)}$ and $t_{i,j}^{(3)}$ are the TOFs of the detected noise photons, and $t_{i,j}^{(2)}$ is the TOF of the detected signal photon.

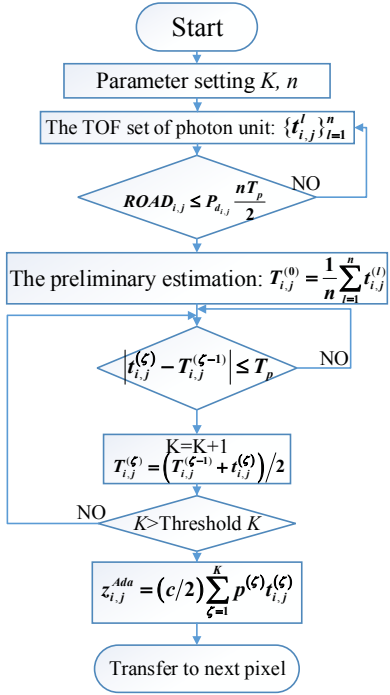

Fig. 2. (b) is the flow diagram of ADI.

### B. Probability of ADI

We define the detection probability of a signal unit ($P(ROAD_{i,j} \leq P_{d_{i,j}} n/2 T_p )$) in the measurement time as $P_w$. $P_w$ indicates the probability that the TOFs of photons converge in the range of $T_p$. Because the width of each time-bin is $T_p$, $P_w$ is converted to the probability that all the photons of a signal unit are distributed in the same time-bin. Assuming that the reflected laser pulse is detected in the $k$-th time-bin (the target time-bin), the probability that the Geiger-mode detector fires in $j$-th ($1 \leq j \leq 800$) time-bin is [17]

$$P_{j<k} = \exp[-(j-1)BT_p][1-\exp(-BT_p)] \quad 1 \leq j < k$$
$$P_{j>k} = \exp[-(j-1)BT_p\text{-}S][1-\exp(-BT_p)] \quad k < j \leq 800 \quad (12)$$
$$P_{j=k} = \exp[-(k-1)BT_P][1-\exp(-S-BT_p)] \quad j = k$$

where $S$ is the mean number of the photon counts generated by the reflected laser light in the measurement time. Since each photon detection is an independent Poisson process [17, 18], the probability that all photons of a signal unit distribute in $j$-th time-bin is $\prod_{l=1}^{n} P_{j<k}^{l}$ when $j < k$, where $n$ is the number of photons in an elementary unit. Thus when $j < k$, the detection probability of a signal unit $P_{w(j<k)}$ is

$$P_{w(j<k)} = \sum_{j=1}^{k-1} \prod_{l=1}^{n} P_{j<k}^{l} . \quad (13)$$

The probability of $P_w$ when $j > k$ or $j = k$ is similar to $P_{w(j<k)}$. Therefore, $P_w$ is computed as

$$P_w = \sum_{j=1}^{k-1} \prod_{l=1}^{n} P_{j<k}^{l} + \prod_{l=1}^{n} P_{j=k}^{l} + \sum_{j=k+1}^{S} \prod_{l=1}^{n} P_{j>k}^{l} . \quad (14)$$

For the Geiger-mode APD, the target detection probability and the false alarm probability in the measurement time are [18]:

$$P_d = \exp[-(k-1)BT_P][1-\exp(-S-BT_p)] \quad (15)$$
$$P_f = 1 - P_d - \exp(-S - Bt_M) \quad (16)$$

where $\exp(-S - Bt_M)$ is the probability that the detector doesn't fire at all in the measurement time (no photon counts). Therefore, the target detection probability and the false alarm probability in the case of using ADI are

$$P_d^{Ada}\left(\text{all }\{t_{i,j}^{(l)}\}_{l=1}^{n}\text{ are signal photon}\Big|ROAD_{i,j} \leq P_{d_{i,j}} n/2 T_p\right) = \frac{\prod_{l=1}^{n} P_d^l}{P_w} \quad (17)$$

$$P_f^{Ada}\left(\{t_{i,j}^{(l)}\}_{l=1}^{n}\text{ at least have a noise photon}\Big|ROAD_{i,j} \leq P_{d_{i,j}} n/2 T_p\right) = \frac{\sum_{l=1}^{n}\binom{l}{n} P_d^{n-l} P_f^l}{P_w} . \quad (18)$$

Conventionally, the maximum likelihood depth estimate (ML estimate) uses the same strategy for the processing of signal photons and noise photons. ML estimate obtains the depth estimation of target using a fixed dwell time in each pixel. Since the photon counts change with the ambient light and the reflectivity of target surface, the recovery accuracy of ML estimate is uncontrollable. However, ADI rejects noise photons, and only signal photons contribute to depth estimation. For all the photon counts are generated by the reflected laser light, the recovery accuracy of ADI is stable. Comparing to the conventional depth estimate methods, ADI takes different dwell time in different area of the target surface. In the case of using ADI, the laser source transmits less laser pulse to the high reflectivity area and more laser pulse to the low reflectivity area. Thus according to the different characteristics of the target surface, ADI can adaptively change the dwell time in each pixel and then obtains the accurate depth image in a short acquisition time.

## III. EXPERIMENT RESULTS

The experimental scene is shown in Figure 3. The target is placed at about 20m distance from the detector. Figure 3(a) is the normalized photon counts generated by the reflected laser light, which corresponds to the different locations in Figure 3(b). The size of target image is $256 \times 256$ pixels in this experiment.

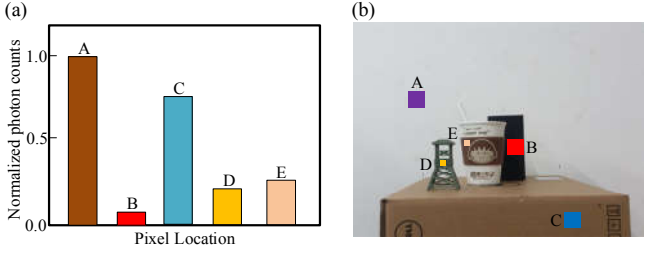

Fig. 3. The experimental target scene. (a) is the normalized photon counts at different locations of the target. (b) is the target scene. The area B is a carton covered with a piece of black cloth, and the area A is a white wall.

The experiment results of ML estimate are shown in Figure 4. The first column from upward to downward is the 3D depth image of target respectively corresponding to the dwell time of 0.5ms, 5ms, 10ms, and 20ms. The second column (high reflectivity area) and third column (low reflectivity area) respectively correspond to the depth estimation of details in area 1 and 2 of the first column.

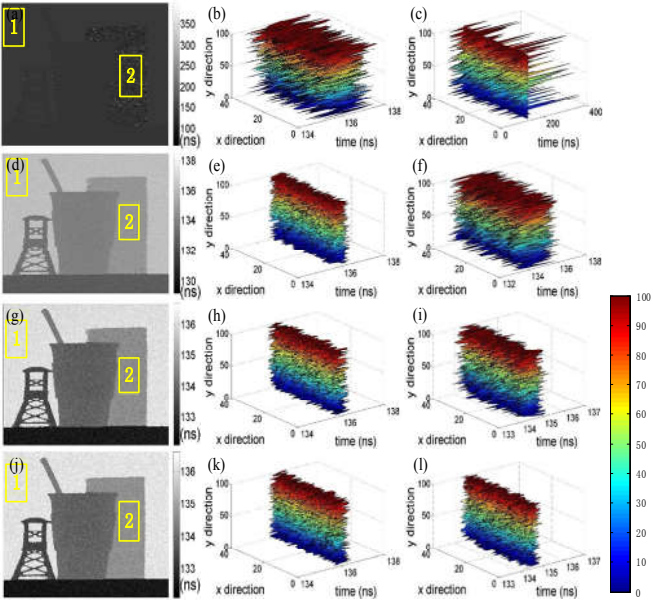

Fig. 4. The experiment results of ML estimate.

Figures 4(a) ~ (c) show that ML estimate cannot accurately estimates the depth of target in a short dwell time. With the dwell time increasing, the improvement of recovery accuracy in the disparate area is different. Even if the accuracy of high reflectivity area is improved shown in Figures 4(a) ~ (c) and Figures 4(d) ~ (f), the depth of low reflectivity area still cannot be accurately estimated. As a result, it results in the problem of under-sampling, as shown in Figure 4(f). With the dwell time increasing, the recovery accuracy of low reflectivity area is gradually improved, as shown in Figures 4(c), (f), (i), (l). However, for the target area with high reflectivity, it's recovery accuracy cannot be improved even at the expense of long dwell time, as shown in Figures 4 (b), (e), (h), (k).

The experiment results of ADI are shown in Figure 5. The adjacent three photons is selected as an elementary unit to estimate the depth of target ($n = 3$). The threshold $K$ is set to 10. Figure 5(a) is the depth image of target. Figure 5(b) is the distribution image of dwell time. Figures 5(c) and 5(d) respectively correspond to the depth detail estimation of area 1 and 2 shown in Figure 5(a). Comparing with Figure 4, it is shown in Figure 5 that the recovery accuracy of ADI is almost unaffected by the target surface reflectivity. Figure 5(b) shows that ADI can adaptively change the dwell time in each pixel according to the target surface reflectivity. Ultimately, the problem of under-sampling or long acquisition time occurred in ML estimate is avoided.

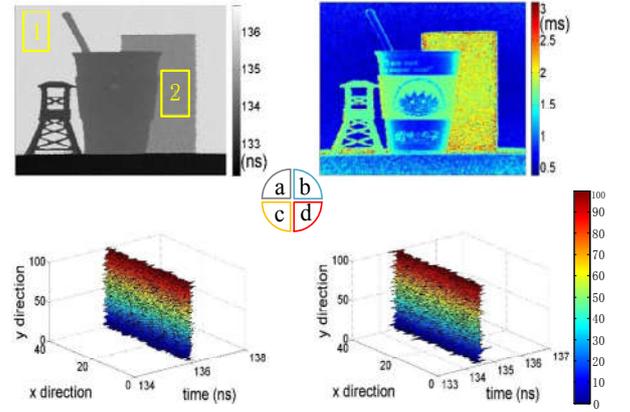

Fig. 5. The experimental results of ADI. The parameters are set that $K$=10 and $n = 3$. (a) is the depth image of target. (b) is the distribution image of dwell time. (c) and (d) correspond to the depth details of area 1 (the white wall, high reflectivity object) and area 2 (the black carton, low reflectivity object) in (a).

Compared with the results of using ML estimate, as shown in Table 1, the RMSE of using ADI is decreased from 1.0392m to 0.0145m, respectively in the condition of the average dwell time of 1.5ms and 1.1ms. At the same time, our method achieves better recovery accuracy than that of ML estimate by reducing the average dwell time from 20ms to 1.1017ms.

Table 1. The comparison of the average dwell time $T_{dwell}$ and the RMSE of depth estimation.

| Imaging method | RMSE/m | $T_{dwell}$/ms |
|---|---|---|
| ML estimate | 1.0392 | 1.5 |
|  | 0.0423 | 20 |
| ADI | 0.0145 | 1.1017 |

At a distance of 20m, the target detection probability at a fixed point of the white wall is acquired with 100 transmitted laser pulses. The parameter $K$ is set to 1 and $n$ is set to 3. This experiment is conducted in different background noise of $SBR = 1$ and $SBR = 10$. The target detection probability of ADI and ML estimate are compared at the same dwell time shown in Table 2.

Table 2. The comparison of the target detection probability

|  | $SBR = 1$ | | $SBR = 10$ | |
|---|---|---|---|---|
|  | ML estimate | ADI | ML estimate | ADI |
| $n = 2$ | 57% | 93% | 85% | 96% |
| $n = 3$ | 58% | 94% | 82% | 99% |
| $n = 4$ | 57% | 97% | 86% | 99% |

As shown in Table 2, the target detection probability in the case of using ADI is increased by nearly 1-fold under the background condition of $SBR = 1$. We find that the size of signal unit in ADI has an effect on the target detection probability: with the size of signal unit increasing, the target detection probability is increased at the same time.

The effect of photon unit size $n$ and background noise on our method are respectively shown in Figure 6 and Table 3. In this experiment, the average TOF of initial detected signal unit $T_{i,j}^{(0)}$ is selected to recover the depth image of target. The experiment results show that with the size of photon unit increasing, the RMSE is improved, but the average dwell time is increased.

Table 3. The effect of photon unit size $n$ and background noise

|  | $SBR = 1$ | | $SBR = 10$ | |
|---|---|---|---|---|
|  | RMSE/m | $T_{dwell}$ /ms | RMSE/m | $T_{dwell}$ /ms |
| $n = 2$ | 0.3474 | 0.1776 | 0.0381 | 0.0440 |
| $n = 3$ | 0.0285 | 0.4233 | 0.0255 | 0.1611 |
| $n = 4$ | 0.02322 | 1.0729 | 0.0219 | 0.3741 |

Figure 7 shows how the performance of ADI is affected by changing the threshold $K$. The target is a carton with black characters of "DELL" in the flat surface, which is placed in 20m away from the detector. In this experiment, the photon unit size $n = 3$ is selected to estimate the depth of target. From the upward to downward, they are respectively the depth image of the flat surface of the carton (left column) and the distribution image of dwell time (right column) corresponding to the threshold of 1, 5, 10 and 20. The experiment results reveal that with the value of the threshold $K$ increasing, the RMSE of using ADI decreases quickly, while the corresponding average dwell time is increased.

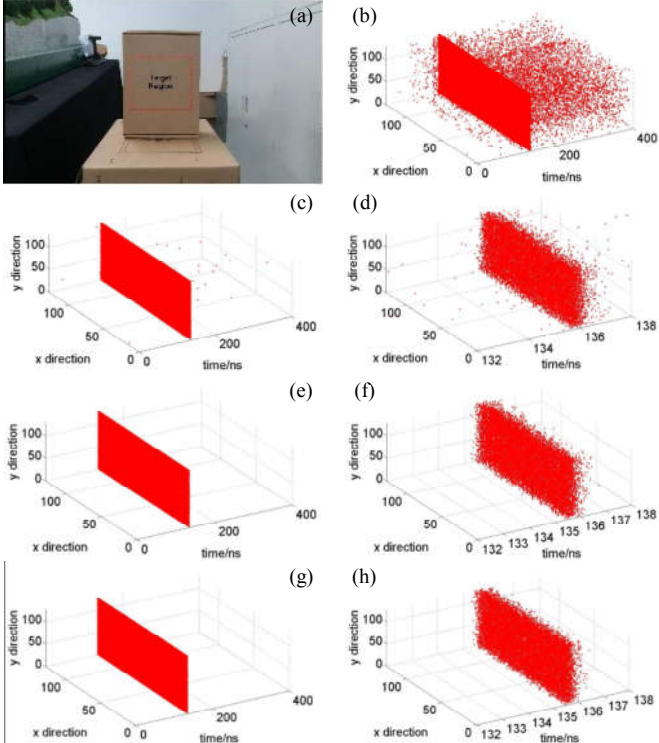

Fig. 6. The effect of the photon unit size $n$. (a) is the experimental target scene. (b) is the original noisy range data. (c), (e), and (g) are the depth image of target respectively corresponding to the experimental results of photon unit size 2, 3, and 4. (d), (f), and (h) are the details of (c), (e), and (f).

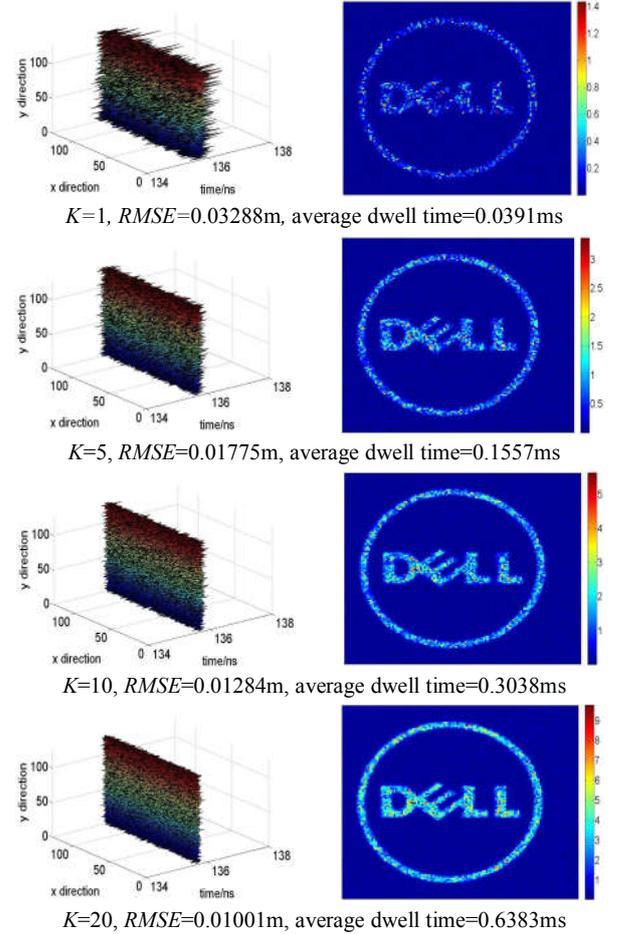

$K=1$, *RMSE*=0.03288m, average dwell time=0.0391ms

$K=5$, *RMSE*=0.01775m, average dwell time=0.1557ms

$K=10$, *RMSE*=0.01284m, average dwell time=0.3038ms

$K=20$, *RMSE*=0.01001m, average dwell time=0.6383ms

Fig. 7. The effect of the threshold $K$. From the upward to downward, they are respectively the depth image of the flat surface of the carton (left column) and the distribution image of dwell time (right column) corresponding to the threshold of 1, 5, 10, and 20.

## IV. CONCLUSIONS

We propose a novel depth imaging strategy that can adaptively change the dwell time in each pixel and fast accurately recover the scene depth. The TOFs of signal photons are mainly and correlatively distributed in the range of the FWHM of the emitted laser pulse; on the other hand, the TOFs of noise photons are uniformly and independently distributed in the pulse repetition period $[0, T_r)$. Using this distribution feature, our method efficiently filters out noise photons, and only signal photons contribute to the depth estimation. According to the characteristics of the target surface, including the surface reflectivity, the method we proposed can adaptively change the dwell time and decrease the useless dwell time in each pixel, which will significantly shorten the acquisition time of depth estimation. In this paper, the operation flow of our method is elaborated, and several experiments are conducted to demonstrate the performance of this method.

Compared with the typical depth estimate method based on maximum likelihood (ML estimate), our method need not to generate the photon-count histogram and has a better performance in recovery accuracy as well as the acquisition time. The experimental results reveal that: the RMSE is improved from 1.0392 m (in ML estimate) to 0.0145 m (in ADI) in an equal average dwell time; in the case of acquiring almost the same accuracy, the average dwell time is decreased from 20 ms (in ML estimate) to 1.1017 ms (in ADI); under the background condition of $SBR=1$, the target detection probability in the case of using ADI is increased by nearly 1-fold in contrast with that of ML estimate. Therefore, our method successfully deals with the problem of under-sampling or long acquisition time in ML estimate and is capable of adaptively taking a clear 3D image, which is useful in rapid or power-limited active optical imaging.

ACKNOWLEDGMENT

The authors gratefully acknowledge the support from the Seventh Six-talent Peak project of Jiangsu Province (Grant Nos. 2014-DZXX-007), the National Natural Science Foundation of China (Grant Nos. 61271332), the Fundamental Research Funds for the Central Universities (Grant Nos. 30920140112012), the Innovation Fund Project for Key Laboratory of Intelligent Perception and Systems for High-Dimensional Information of Ministry of Education (Grant Nos. JYB201509), and the Fund Project for Low-light-level Night Vision Laboratory (Grant Nos. J20130501).